\documentclass[9pt,conference]{IEEEtran}

\usepackage{graphicx}
\DeclareGraphicsExtensions{.pdf,.png,.jpg}

\usepackage{pgfplots} 
\pgfplotsset{compat=newest} 
\pgfplotsset{plot coordinates/math parser=false} 
\usetikzlibrary{plotmarks}
\usetikzlibrary{positioning, calc}
\usepackage{grffile}
\newlength\figureheight 
\newlength\figurewidth

\usepackage{amsmath}
\usepackage{amssymb}
\usepackage{amsthm}

\usepackage{cite}
\usepackage{setspace}

\usepackage{dsfont}
\usepackage{mathbbol}
\usepackage{bm}
\usepackage{bbm}

\usepackage{array}
\usepackage{mdwmath}
\usepackage{mdwtab}
\usepackage{eqparbox}

\usepackage{overpic}

\usepackage{algorithm}
\usepackage{algorithmic}

\usepackage{color}

\usepackage{listings}

\newcommand{\T}{\top}
\newcommand{\mym}{\mathbf{m}}
\newcommand{\myd}{\mathbf{d}}
\newcommand{\myb}{\mathbf{b}}
\newcommand{\myA}{\mathbf{A}}
\newcommand{\myx}{\mathbf{x}}
\newcommand{\N}{\mathcal{N}}

\IEEEoverridecommandlockouts

\begin{document}

\title{Simulation of Underwater RF Wireless Sensor Networks using Castalia}

\author{
	\IEEEauthorblockN{
		Sergio Valcarcel Macua, 
		Santiago Zazo\\
		Javier Zazo, Marina P\'erez Jim\'enez 
		\thanks{
			Work partially supported  by the Spanish Ministry of Science
			and Innovation grant TEC2013-46011-C3-1-R, 
			the COMONSENS Network of Excellence TEC2015-69648-REDC 
			and by an FPU doctoral grant to Javier Zazo.
		}
	}
	\IEEEauthorblockA{
		Universidad Polit\'ecnica de Madrid,\\
		Madrid 28040, Spain.\\
		Email: \{sergio, santiago\}@gaps.ssr.upm.es,\\
		javier.zazo.ruiz@upm.es,
		marina.perez@isom.upm.es
	}
	\and
	\IEEEauthorblockN{
		Iv\'an  P\'erez-\'Alvarez, 
		Eugenio Jim\'enez
	}
	\IEEEauthorblockA{
		Instituto para el Desarrollo Tecnol\'ogico y la\\
		Innovaci\'on en Comunicaciones(IDeTIC),\\
		Universidad de Las Palmas de Gran Canaria,\\
		Las Palmas, 35017, Spain.\\
		Email: \{ivan.perez, eugenio.jimenez\}@ulpgc.es
	}
	\and
	\IEEEauthorblockN{
		Joaqu\'in Hern\'andez Brito
	}
	\IEEEauthorblockA{
		Plataforma Oce\'anica de\\
		Canarias (PLOCAN),\\
		Telde 35200, Spain \\
		Email: joaquin.brito@plocan.eu
		}
}


\maketitle

%
\begin{abstract}

We use real measurements of the underwater channel 
to simulate a whole underwater RF wireless sensor networks,
including propagation impairments (e.g., noise, interferences), 
radio hardware (e.g., modulation scheme, bandwidth, transmit power), 
hardware limitations (e.g., clock drift, transmission buffer)
and complete MAC and routing protocols.
The results should be useful for designing centralized and distributed algorithms
for applications like monitoring, event detection, localization and aid to navigation.
We also explain the changes that have to be done to Castalia
in order to perform the simulations.
\end{abstract}
%


\IEEEpeerreviewmaketitle

%
\section{Introduction}
%

Underwanter sensor networks are useful for environmental monitoring and security applications.
While acoustic or optical methods are preferred in deep sea water scenarios,
they suffer from several impairments in shallow water settings.
In particular, acoustic signals suffer from multi-path propagation, reverberation and ambient noise and, very importantly, they have a negative impact on marine life
\cite{ jepson2003gas,parsons2008navy};
while optical signals suffer from high absorption and strong backscatter
\cite{wang2014experimental}.
RF communications seems an attractive alternative that could offer higher bandwidth and better transmission in medium boundaries (e.g., water--air, seabed--ice).
Indeed, we have been able to measure the frequency response and achieve stable links at some meters distance \cite{Mena2016, Jimenez2016}.
Now, we want to develop underwater wireless sensor networks (U-WSN)
for applications like 
localization,
aid to navigation,
event detection and environment monitoring.
We consider both centralized and distributed algorithms.
In a centralized algorithm, 
nodes sense the environment 
and send their measurements to a \textit{sink} node,
which is in charge of gathering and processing all data from every node.
Distributed algorithms are those in which nodes communicate with their neighbors
and exchange some information 
(not necessarily the measurements but some intermediate variables like their estimate about some magnitude),
so that they can predict and interact with the environment.
Examples of distributed algorithms include feature extraction for data compression \cite{Belanovic2012Distri}, 
diffusion adaptation \cite{Sayed2013SPMagazine}
and belief propagation algorithms \cite{Savic2014149}.
From a communications point of view, 
centralized algorithms require routing any packet from any node to the sink,
which requires route discovery;
while distributed algorithms require point to point communication between neighbors,
which may require neighborhood discovery.

Deploying U-WSN presents several challenges,
like communication impairments, unexpected and asynchronous events, 
limited battery life, etc.
Thus, we approach the problem in four main stages:
\textit{i)} channel characterization,
\textit{i)} simulation of U-WSN for specific scenario,
\textit{iii)} algorithm design 
and \textit{iv)} network deployment.
Stage 1 includes measurement campaigns to obtain the 
frequency response and the underwater propagation model underwater. 
In this paper we focus on stage 2,
using the underwater channel characterization explained in 
the companion papers \cite{Mena2016, Jimenez2016}.
In particular,
the main contribution of this work is to simulate a whole U-WSN,
including propagation impairments (e.g., noise, interferences), 
radio hardware (e.g., modulation scheme, bandwidth, transmit power), 
hardware limitations (e.g., clock drift, transmission buffer),
complete MAC and routing protocols
for studying the influence of different parameters on some standard scenarios.
The results of this work should be input to stage 3 for studying the feasibility of some algorithms
(e.g., the order of loss packet rate that must be tolerated by the distributed algorithms presented in the companion paper \cite{Zazo2016}).
Stage 4 includes prototype development and building a testbed.
The idea is to iterate over these stages in order to minimize the cost of a real deployment.

Castalia \cite{pediaditakis2010performance} is a simulator for Wireless Sensor Networks (WSN)
based on the OMNeT++ platform  \cite{Varga2008Omnet}
that offers realistic wireless channel and radio models and realistic node behaviour.
The main reasons for choosing Castalia are its level of realism, speed and flexibility.
The speed is achieved because all the modules are written in C++.
The flexibility is at the cost of having no GUI.
Indeed, Castalia is a command line simulator,
where scenarios and settings are defined through external configuration files.
The results are also given in text files, 
but they can be accessed with convenient parser scripts.
In the following sections we explain how we have used and extended Castalia for simulating a U-WSN in some scenarios.

%
\section{Wireless Channel}
%

%
\subsection{Channel model characterization}
%

\begin{figure}[!ht]
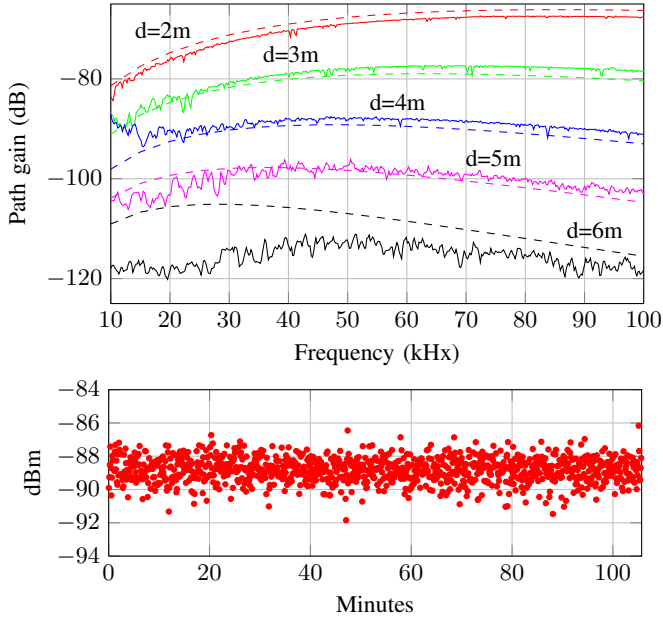

\centering
\input{measurements_2.tikz}
\\
\input{fading.tikz}
\vspace{-0.5em}
\caption{Measurements of the underwatter wireless channel. 
Frequency response at multiple distances (top). 
Temporal variation at 46kHz (bottom).} 
\label{Fig:wireless-channel}
\end{figure}

We start from the measurements of the underwater channel--taken with loop antennas---presented in the companion papers \cite{Mena2016,Jimenez2016}.
Figure \ref{Fig:wireless-channel}-top shows the measured (solid)
and simulated (dashed) channel frequency response.
Figure \ref{Fig:wireless-channel}-bottom shows the fading of the signal at 46kHz along almost 2 hours.

We have used the measurements of the frequency response to propose a path loss model,
where the log of the attenuation is linear with the distance.
Let $L(d)$ denote the path loss (dB) at $d$ meters from the transmitter.
Then, the proposed model expresses $L(d)$ as an affine function of the distance
with parameter $\eta$:
\begin{align}
L(d) 
= 
	L(d_0) + \eta \frac{d}{d_{0}} 
	+
	X
\label{eq:linear-model}
\end{align}
where $d_0$ is some reference distance 
and $X$ is a random variable.

Note that \eqref{eq:linear-model} is different from the standard free-space path loss model,
in which the log of the attenuation is linear with the log of the distance:
\begin{align}
L_{\rm fs}(d) 
= 
	L (d_0) + \eta \cdot 10 \cdot \log \left( \frac{d}{d_{0}} \right)
	+
	X
\label{eq:free-space-model}
\end{align}

We remark that these path loss models should be frequency dependent, 
but we have assumed the simpler narrow band case.

In order to find parameters $L(d_0)$ and $\eta$ for both models,
we have considered carrier frequency $46$kHz
and $d_0 = 2$.
The vector of distances is
$\myd = [2, 3, 4, 5, 6]^{\T}$
and the corresponding losses are 
$\mym = [78.5, 88, 98, 112 ]^{\T}$ 
(note that the loss is opposite sign to the gain displayed in Figure  \ref{Fig:wireless-channel}).
Let 
$\myx =  [  \eta, L(d_0) ]^\T$ be the vector of unknown parameters.
Now, for the linear model, we build a new vector 
$\myb = \myd / d_0$ 
and introduce the matrix $\myA = [\myb, \bm{1}]$ of size $6 \times 2$,
where $\bm{1}$ is a vector of ones.
Thus, we have a system of equations,
$
\mym = \myA \myx
$,
that we can solve as follows:
$
\myx
=
(\myA^\T \myA)^{-1} \myA^\T \mym
$,
obtaining linear coefficient $\eta/d_0 = 10.45$ and offset $L(d_0) = 47.40$dB.
The procedure for finding the parameters of the free-space model is the same as for the linear model but replacing $\myb$ by the scaled $\log$ of the distance, i.e.,
$\myb = 10 \cdot \log (\myd / d_0)$.

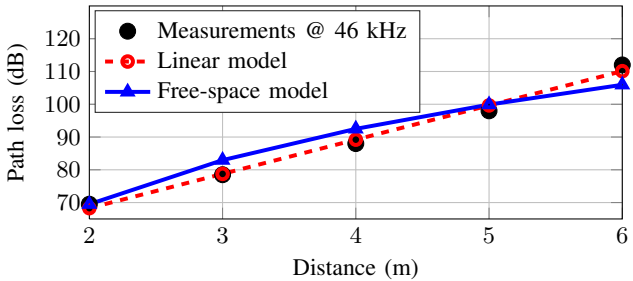
\begin{figure}[!ht]
\centering
%
%
\begin{tikzpicture}

\begin{axis}[%
width=0.8\columnwidth,
height=0.32\columnwidth,
scale only axis,
xmin=2,
xmax=6,
xtick={2, 3, 4, 5, 6},
xlabel={Distance (m)},
xmajorgrids,
ymin=65,
ymax=130,
ytick={ 70,  80,  90, 100, 110, 120},
ylabel={Path loss (dB)},
ymajorgrids,
axis background/.style={fill=white},
legend style={at={(0.01,0.98)},anchor=north west,legend cell align=left,align=left,draw=white!15!black}
]
\addplot [color=black,mark size=3.0pt,only marks,mark=*,mark options={solid,fill=black,draw=black}]
  table[row sep=crcr]{%
2	69.5\\
3	78.5\\
4	88\\
5	98\\
6	112\\
};
\addlegendentry{Measurements @ 46 kHz};

\addplot [color=red,dashed,line width=1.5pt,mark=o,mark options={solid,fill=red,draw=red}]
  table[row sep=crcr]{%
2	68.3\\
3	78.75\\
4	89.2\\
5	99.65\\
6	110.1\\
};
\addlegendentry{Linear model};

\addplot [color=blue,solid,line width=1.5pt,mark=triangle,mark options={solid,draw=blue}]
  table[row sep=crcr]{%
2	69.5\\
3	82.949913882598\\
4	92.492779650006\\
5	99.8948013988973\\
6	105.942693532604\\
};
\addlegendentry{Free-space model};

\end{axis}
\end{tikzpicture}%
\vspace{-0.5em}
\caption{Measurements and path loss models.} 
\label{Fig:channel_coefficient}
\end{figure}

Figure \ref{Fig:channel_coefficient} shows the real measurements at $46$kHz (black dots),
the proposed linear model (dashed red) and the free-space model (solid blue).
It is apparent that the proposed linear model \eqref{eq:linear-model} fits the measurements 
much better than the free-space model \eqref{eq:free-space-model}.

\begin{figure}[!ht]
\centering
%
%
\begin{tikzpicture}

\begin{axis}[%
width=0.8\columnwidth,
height=0.2\columnwidth,
scale only axis,
xmin=-92,
xmax=-86,
ymin=0,
ymax=200,
xlabel=dBm,
axis background/.style={fill=white}
]
\addplot[ybar,bar width=0.3,draw=black,fill=white!80!blue,area legend] plot table[row sep=crcr] {%
-91.85	1\\
-91.55	1\\
-91.25	1\\
-90.95	7\\
-90.65	11\\
-90.35	24\\
-90.05	41\\
-89.75	56\\
-89.45	119\\
-89.15	160\\
-88.85	160\\
-88.55	185\\
-88.25	121\\
-87.95	86\\
-87.65	49\\
-87.35	23\\
-87.05	7\\
-86.75	3\\
-86.45	1\\
-86.15	1\\
};
\addplot [color=red,solid,line width=2.0pt,forget plot]
  table[row sep=crcr]{%
-91.0736123159864	1.87687997583538\\
-91.0282328296838	2.24699040306751\\
-90.9828533433812	2.68022184114428\\
-90.9374738570786	3.18526136608225\\
-90.892094370776	3.77158772844494\\
-90.8467148844734	4.44946875774825\\
-90.8013353981709	5.22994253655051\\
-90.7559559118683	6.1247798127319\\
-90.7105764255657	7.14642518159902\\
-90.6651969392631	8.30791472234842\\
-90.6198174529605	9.62276802651293\\
-90.574437966658	11.104852917602\\
-90.5290584803554	12.7682216366158\\
-90.4836789940528	14.6269178593318\\
-90.4382995077502	16.6947546157442\\
-90.3929200214476	18.9850639925416\\
-90.347540535145	21.5104204034542\\
-90.3021610488425	24.2823401916443\\
-90.2567815625399	27.3109613593233\\
-90.2114020762373	30.6047082735506\\
-90.1660225899347	34.1699472397446\\
-90.1206431036321	38.0106398279747\\
-90.0752636173296	42.128001740679\\
-90.029884131027	46.5201757816396\\
-89.9845046447244	51.1819280822812\\
-89.9391251584218	56.1043771221306\\
-89.8937456721192	61.2747652085266\\
-89.8483661858166	66.6762819253256\\
-89.8029866995141	72.2879485977757\\
-89.7576072132115	78.0845720368839\\
-89.7122277269089	84.0367747181076\\
-89.6668482406063	90.111107125138\\
-89.6214687543037	96.2702462713815\\
-89.5760892680012	102.473282434278\\
-89.5307097816986	108.67609394801\\
-89.485330295396	114.831807557451\\
-89.4399508090934	120.891339409113\\
-89.3945713227908	126.804009320434\\
-89.3491918364882	132.518218608783\\
-89.3038123501857	137.982179560344\\
-89.2584328638831	143.144682659518\\
-89.2130533775805	147.955886059836\\
-89.1676738912779	152.368110527878\\
-89.1222944049753	156.336622290975\\
-89.0769149186728	159.820385911936\\
-89.0315354323702	162.782769527567\\
-88.9861559460676	165.192185531541\\
-88.940776459765	167.022651046289\\
-88.8953969734624	168.254254283568\\
-88.8500174871599	168.873515090761\\
-88.8046380008573	168.87363055457\\
-88.7592585145547	168.25459940501\\
-88.7138790282521	167.023222038662\\
-88.6684995419495	165.192976160588\\
-88.623120055647	162.783771224783\\
-88.5777405693444	159.821587929134\\
-88.5323610830418	156.338011892221\\
-88.4869815967392	152.369673214003\\
-88.4416021104366	147.957605819982\\
-88.396222624134	143.146542243642\\
-88.3508431378315	137.984160766192\\
-88.3054636515289	132.520302576697\\
-88.2600841652263	126.806176829356\\
-88.2147046789237	120.89357116737\\
-88.1693251926211	114.834084482278\\
-88.1239457063186	108.678397428323\\
-88.078566220016	102.475594571872\\
-88.0331867337134	96.2725500965152\\
-87.9878072474108	90.1133867833113\\
-87.9424277611082	84.039015625769\\
-87.8970482748056	78.0867610046498\\
-87.8516687885031	72.2900739207073\\
-87.8062893022005	66.6783334406109\\
-87.7609098158979	61.2767343217098\\
-87.7155303295953	56.1062568037218\\
-87.6701508432927	51.1837128371619\\
-87.6247713569902	46.5218615941544\\
-87.5793918706876	42.1295859986577\\
-87.534012384385	38.0121212291556\\
-87.4886328980824	34.1713256841896\\
-87.4432534117798	30.6059847456417\\
-87.3978739254772	27.3121378028349\\
-87.3524944391747	24.2834193812019\\
-87.3071149528721	21.5114058154313\\
-87.2617354665695	18.9859596781342\\
-87.2163559802669	16.6955650786048\\
-87.1709764939643	14.6276479397647\\
-87.1255970076618	12.768876403942\\
-87.0802175213592	11.1054375719126\\
-87.0348380350566	9.62328781065416\\
-86.989458548754	8.30837484462823\\
-86.9440790624514	7.14683074938756\\
-86.8986995761488	6.12513577684309\\
-86.8533200898463	5.23025364605177\\
-86.8079406035437	4.44973952464663\\
-86.7625611172411	3.77182240154211\\
-86.7171816309385	3.1854639131769\\
-86.6718021446359	2.68039593868641\\
-86.6264226583334	2.24713943232629\\
-86.5810431720308	1.87700702463967\\
};
\end{axis}
\end{tikzpicture}%
\vspace{-0.5em}
\caption{Histogram and normal distribution fit 
for the trace of measurements.} 
\label{Fig:random-variable}
\end{figure}
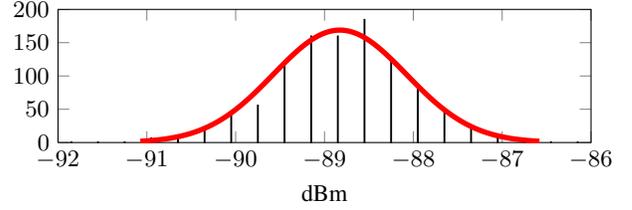

The additive noise $X$ can be also estimated from the trace of measurements displayed in Figure \ref{Fig:wireless-channel}-bottom.
Figure \ref{Fig:random-variable} shows the histogram for such trace
together with a normal distribution fit.
We conclude that the Normal distribution is a reasonable approximation, 
so that we can assume $X \sim \N ( 0, \sigma^2)$,
with $\sigma^2 = 0.56$dB.

A final observation from Figure \ref{Fig:wireless-channel}-bottom relevant to the wireless channel simulation is that we can dismiss slow fading.

%
\subsection{Simulation of the linear channel model}
%

Castalia offers only one parametric channel out of the box, 
which is the free-space model.
Nevertheless, it also offers the possibility of expressing non-parametric path loss models by 
describing the path loss for every pair of nodes in an external file.
For instance, suppose we have a network of $N$ nodes, 
then, for every node $n$,
we can express the path loss seen by every node $m \ne n$ when node $n$ is transmitting.
Although flexible, 
this approach requires to rewrite the path loss file every time we change the network topology or the model parameters.
Hence, we preferred to extend Castalia by including the linear model \eqref{eq:linear-model}
into the class \textbf{WirelessChannel}.

For the temporal variation of the channel gain, 
Castalia only offers the possibility 
of writing an external file describing the fading probability distribution 
at every time instant, allowing to include temporal correlation.
Again, although flexible, 
the main handicap of using a external files is that we have to rewrite them every time we change a parameter (e.g., the variance of the PDF).
In our particular case, 
Figure \ref{Fig:random-variable}
shows that a simple additive Gaussian random variable fits well the measurements.
Hence, we have found more convenient to extend Castalia by adding a normally distributed fast fading model to the class \textbf{channelTemporalModel}.

%

%
\section{Radio Parameters}
%

We have chosen an FSK modulation scheme since it is optimal in the low signal-to-noise-ratio (SNR) regime.
Castalia offers FSK with noncoherent receiver and no channel coding.
We have added coherent receiver and channel coding to the class 
\textbf{Radio} (in particular, to the method \textbf{Radio::SNR2BER}).
We have selected a convolutional code $(3,1,2)$,
whose nonsystematic feedforward encoder is represented by 
$G(D) = [1+D, \: 1+D^2, \: 1 + D + D^2]$,
with free distance $d_{\rm free}=7$ 
and $B_{d_{\rm free}} = 1$
(this is the total number of nonzero information bits on all weight-$d_{\rm free}$ paths,
divided by the number of information bits per unit time).
Let $p$ be the BER obtained from the SNR curve for our FSK modulation.
We have characterized the system by including the coding gain as the BER upper bound given by \cite[Eq. 12.1]{lin2004error}:
\begin{align}
{\rm BER} (p, B_{d_{\rm free}},d_{\rm free}) 
\approx 
	B_{d_{\rm free}}
	\left[
		2 \sqrt{p (1-p)}
	\right]^{d_{\rm free}}
\end{align}

The radio parameters common to all simulations are:
data rate $3$kbps,
$1$ bit per symbol,
signal bandwidth $12$kHz,
noise bandwidth $12$kHz, 
noise floor $-108$dBm
and  
receiver sensitivity $-101$dBm.

The required transmit power depends on the scenario under study
and influences critically on the communication range. 
Section \ref{ssec:point-to-point} shows simulation results for $10$dBm, $20$dBm and $30$dBm,
while Sections \ref{ssec:routing-ack}--\ref{ssec:diffusion-broadcast}
use $30$dBm.
Castalia has a prefixed value of $0$dBm as the maximum power allowed by the simulator,
which must be updated for our setting. 
In particular, it is needed to increase the value of variable \textbf{maxTxPower} 
in class \textbf{WirelessChannel}.

The maximum packet size depends on the application under study.
For instance, 
the localization algorithm considered in the companion paper \cite{Zazo2016}
requires small packets of size $8$ bytes (i.e., $2$ float numbers per iteration) 
before encoding.
Recall that the larger the packet, the higher the probability that it suffers some impairment.
Section \ref{ssec:point-to-point} shows simulation results for random packets of $30$ bytes,
while the rest of simulations transmit their neighbor list, 
which consists of around $15$ bytes.

%
\section{Scenarios under Test}
%

%
\subsection{Point to point}
\label{ssec:point-to-point}
%

\begin{figure}[!ht]
\centering
\includegraphics[width=0.55\columnwidth]{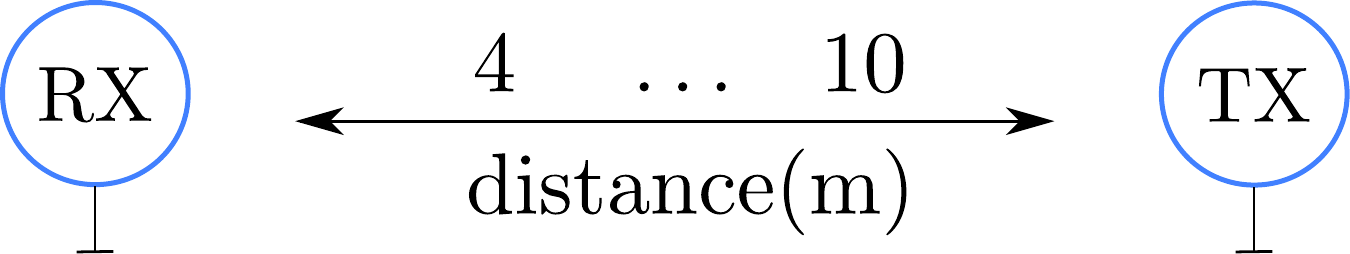}
\vspace{-0.5em}
\caption{Point to point scenario with 2 nodes.} 
\label{Fig:scenario-p2p}
\end{figure}

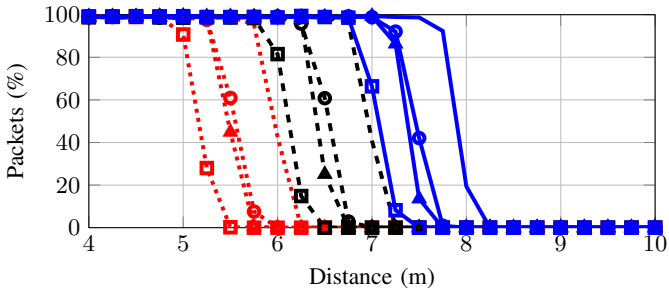
\begin{figure}[!ht]
\centering
%
%
\begin{tikzpicture}

\begin{axis}[%
width=0.85\columnwidth,
height=0.32\columnwidth,
scale only axis,
xmin=4,
xmax=10,
ymin=0,
ymax=100,
xmajorgrids,
ymajorgrids,
xlabel=Distance (m),
ylabel=Packets (\%),
axis background/.style={fill=white},
legend style={at={(0.98,0.98)},anchor=north east,legend cell align=left,align=left,draw=white!15!black}
]
\addplot [color=red,dotted,line width=1.5pt,mark=square,mark options={solid,fill=red,draw=red}]
  table[row sep=crcr]{%
4	98.9\\
4.25		99\\
4.5	98.9\\
4.75		99\\
5	90.686\\
5.25	 28.004\\
5.5	0.2\\
5.75		0\\
6		0\\
6.25		0\\
6.5	0\\
6.75		0\\
7	0\\
7.25		0\\
7.5	0\\
7.75		0\\
8	0\\
8.25		0\\
8.5	0\\
8.75		0\\
9	0\\
9.25		0\\
9.5	0\\
9.75		0\\
10	0\\
};

\addplot [color=red,dotted,line width=1.5pt,mark=o,mark options={solid,fill=red,draw=red}]
  table[row sep=crcr]{%
4	99.1\\
4.25		99.2\\
4.5	99.2\\
4.75		99.4\\
5	99\\
5.25		97.584\\
5.5	60.916\\
5.75		7.584\\
6	0\\
6.25		0\\
6.5	0\\
6.75		0\\
7	0\\
7.25		0\\
7.5	0\\
7.75		0\\
8	0\\
8.25		0\\
8.5	0\\
8.75		0\\
9	0\\
9.25		0\\
9.5	0\\
9.75		0\\
10	0\\
};
\addplot [color=red,dotted,line width=1.5pt,mark=triangle,mark options={solid,fill=red,draw=red}]
  table[row sep=crcr]{%
4	98.9\\
4.25		98.9\\
4.5	99.1\\
4.75		98.8\\
5	98.4\\
5.25		98.09\\
5.5	44.871\\
5.75		0.508\\
6	0\\
6.25	0\\
6.5	0\\
6.75	0\\
7	0\\
7.25	0\\
7.5	0\\
7.75	0\\
8	0\\
8.25	0\\
8.5	0\\
8.75	0\\
9	0\\
9.25	0\\
9.5	0\\
9.75	0\\
10	0\\
};
\addplot [color=red,dotted,line width=1.5pt]
  table[row sep=crcr]{%
4	99\\
4.25		98.9\\
4.5	99\\
4.75		99.1\\
5	98.8\\
5.25		98.7\\
5.5	99\\
5.75		95.386\\
6	42.594\\
6.25		0.204\\
6.5	0\\
6.75	0\\
7	0\\
7.25	0\\
7.5	0\\
7.75	0\\
8	0\\
8.25	0\\
8.5	0\\
8.75	0\\
9	0\\
9.25	0\\
9.5	0\\
9.75	0\\
10	0\\
};
\addplot [color=black,dashed,line width=1.5pt,mark=square,mark options={solid,fill=black,draw=black}]
  table[row sep=crcr]{%
4	98.8\\
4.25		99.1\\
4.5	99.3\\
4.75		98.7\\
5	98.6\\
5.25		98.9\\
5.5	98.7\\
5.75		98.4\\
6	81.484\\
6.25		14.843\\
6.5	0\\
6.75	0\\
7	0\\
7.25	0\\
7.5	0\\
7.75	0\\
8	0\\
8.25	0\\
8.5	0\\
8.75	0\\
9	0\\
9.25	0\\
9.5	0\\
9.75	0\\
10	0\\
};
\addplot [color=black,dashed,line width=1.5pt,mark=o,mark options={solid,fill=black,draw=black}]
  table[row sep=crcr]{%
4	99.2\\
4.25		99.3\\
4.5	99.3\\
4.75		98.6\\
5	99.1\\
5.25		98.9\\
5.5	98.8\\
5.75		98.9\\
6	99.2\\
6.25		95.788\\
6.5	60.798\\
6.75		2.922\\
7	0\\
7.25	0\\
7.5	0\\
7.75	0\\
8	0\\
8.25	0\\
8.5	0\\
8.75	0\\
9	0\\
9.25	0\\
9.5	0\\
9.75	0\\
10	0\\
};
\addplot [color=black,dashed,line width=1.5pt,mark=triangle,mark options={solid,fill=black,draw=black}]
  table[row sep=crcr]{%
4	98.7\\
4.25		99.1\\
4.5	99\\
4.75		99\\
5	99.1\\
5.25		98.6\\
5.5	99.1\\
5.75		99\\
6	98.8\\
6.25		97.694\\
6.5	25.086\\
6.75		0.102\\
7	0\\
7.25	0\\
7.5	0\\
7.75	0\\
8	0\\
8.25	0\\
8.5	0\\
8.75	0\\
9	0\\
9.25	0\\
9.5	0\\
9.75	0\\
10	0\\
};
\addplot [color=black,dashed,line width=1.5pt]
  table[row sep=crcr]{%
4	99\\
4.25		98.5\\
4.5	99.1\\
4.75		98.9\\
5	99.2\\
5.25		99\\
5.5	99\\
5.75		99.2\\
6	99\\
6.25		99\\
6.5	98.9\\
6.75		96.288\\
7	40.786\\
7.25	0\\
7.5	0\\
7.75	0\\
8	0\\
8.25	0\\
8.5	0\\
8.75	0\\
9	0\\
9.25	0\\
9.5	0\\
9.75	0\\
10	0\\
};
\addplot [color=blue,solid,line width=1.5pt,mark=square,mark options={solid,fill=blue,draw=blue}]
  table[row sep=crcr]{%
4	99.2\\
4.25		99\\
4.5	99.2\\
4.75		99.2\\
5	99.3\\
5.25		98.8\\
5.5	98.9\\
5.75		99\\
6	98.7\\
6.25		99.4\\
6.5	98.9\\
6.75		98.598\\
7	66.353\\
7.25		8.184\\
7.5	0\\
7.75	0\\
8	0\\
8.25	0\\
8.5	0\\
8.75	0\\
9	0\\
9.25	0\\
9.5	0\\
9.75	0\\
10	0\\
};
\addplot [color=blue,solid,line width=1.5pt,mark=o,mark options={solid,fill=blue,draw=blue}]
  table[row sep=crcr]{%
4	98.8\\
4.25		98.9\\
4.5	99\\
4.75		98.9\\
5	99.2\\
5.25		98.9\\
5.5	99\\
5.75		99\\
6	98.9\\
6.25		99\\
6.5	99\\
6.75		99.2\\
7	98.8\\
7.25		92.053\\
7.5	42.069\\
7.75		0.912\\
8	0\\
8.25	0\\
8.5	0\\
8.75	0\\
9	0\\
9.25	0\\
9.5	0\\
9.75	0\\
10	0\\
};
\addplot [color=blue,solid,line width=1.5pt,mark=triangle,mark options={solid,fill=blue,draw=blue}]
  table[row sep=crcr]{%
4	99.1\\
4.25		98.6\\
4.5	98.9\\
4.75		99.2\\
5	99.3\\
5.25		99.1\\
5.5	98.7\\
5.75		98.7\\
6	99\\
6.25		98.9\\
6.5	98.9\\
6.75		98.7\\
7	99.3\\
7.25		86.288\\
7.5	13.571\\
7.75		0\\
8	0\\
8.25	0\\
8.5	0\\
8.75	0\\
9	0\\
9.25	0\\
9.5	0\\
9.75	0\\
10	0\\
};
\addplot [color=blue,solid,line width=1.5pt]
  table[row sep=crcr]{%
4	99.1\\
4.25		98.7\\
4.5	98.9\\
4.75		98.9\\
5	99\\
5.25		99.3\\
5.5	99\\
5.75		99.1\\
6	99\\
6.25		99\\
6.5	98.5\\
6.75		99.1\\
7	99.3\\
7.25		98.9\\
7.5	98.7\\
7.75		92.369\\
8	19.398\\
8.25	0\\
8.5	0\\
8.75	0\\
9	0\\
9.25	0\\
9.5	0\\
9.75	0\\
10	0\\
};
\end{axis}
\end{tikzpicture}%
\vspace{-0.5em}
\caption{Successfully received packets vs. distance (m).} 
\label{Fig:p2p-packet-rate}
\end{figure}

The first scenario under study is a 2-nodes, point-to-point setting, 
where one node transmits and the other listens
(see Figure \ref{Fig:scenario-p2p}). 
We study the successfully received packet rate for different distances between $4$ and $10$ meters.
Figure \ref{Fig:p2p-packet-rate} shows results (averaged over 20 independent trials) 
for 3 different transmit power:
$10$dBm (red dotted),
$20$dBm (black dashed)
and 
$30$dBm (blue solid);
and for 4 different receivers:
noncoherent with no channel coding (square),
coherent with no channel coding (circle),
noncoherent with channel coding (triangle)
and
coherent with channel coding (no marker).
As expected, including a coherent receiver and some channel coding improves the rate.
We conclude that with the current radio system, 
the maximum reliable distance is around $4.75$m for $10$dBm,
$6.75$m for $20$dBm
and $7.5$m for $30$dBm.

%
\subsection{Network routing using unicast transmissions with ACK}
\label{ssec:routing-ack}
%

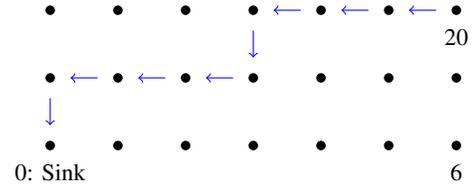
\begin{figure}[!ht]
\centering
\tikzset{
circle node/.style = {circle,inner sep=1pt,draw, fill=white},
dot node/.style = {circle, draw, inner sep=5pt},
X node/.style = {fill=white, inner sep=1pt}
}

\begin{tikzpicture}[scale=0.9]
    \foreach \x in {0,...,6}
    \foreach \y in {0,...,2}
    {
    \fill (\x,\y) circle (2pt);
    }
    \node [X node] at (0,-0.4) {0: Sink};
    \node [X node] at (6,-0.4) {6};
    \node [X node] at (6, 2-0.4) {20};

  \draw[-> , blue] (5.7,2) -- (5.3,2);
  \draw[-> , blue] (4.7,2) -- (4.3,2);
  \draw[-> , blue] (3.7,2) -- (3.3,2);
  \draw[-> , blue] (3,1.7) -- (3,1.3);
  \draw[-> , blue] (2.7,1) -- (2.3,1);
  \draw[-> , blue] (1.7,1) -- (1.3,1);
  \draw[-> , blue] (0.7,1) -- (0.3,1);
  \draw[-> , blue] (0,0.7) -- (0,0.3);

\end{tikzpicture}
\vspace{-0.5em}
\caption{Routing scenario.} 
\label{Fig:scenario-routing}
\end{figure}

In this scenario we simulate a complete network of 21 nodes,
which are deployed over a field of $42 \times 18$ square meters
in a 3 rows, 
7 columns grid
(see Figure \ref{Fig:scenario-routing}).
The main application is to collect data from the network by a sink.

We choose the Collection Tree Protocol (CTP) for routing packets to the sink \cite{gnawali2013ctp,FonsecaCTPTep123,colesanti2010performance}.
CTP has been implemented in several operating system 
(e.g., TinyOS, Mantis OS, Contiki OS, Sun SPOTs).
It is also offered by Castalia \cite{colesanti2011collection} as a set of routing and MAC modules.
CTP uses 3 techniques to provide efficient and reliable routing:
\textit{i)} an accurate link estimator that uses feedback from both the data and control planes;
\textit{ii)} the Trickle algorithm \cite{levis2003trickle} to time the control traffic, 
sending few beacons in stable topologies yet quickly adapting to changes;
\textit{iii)} actively probing the topology with data traffic, quickly discovering and fixing routing failures.
CTP relies in acknowledged unicast transmissions for estimating the quality of the links.

\begin{figure}[!ht]
\centering
%
%
\definecolor{mycolor1}{rgb}{0.00000,0.44700,0.74100}%
\begin{tikzpicture}

\begin{axis}[%
width=0.8\columnwidth,
height=0.2\columnwidth,
scale only axis,
xmin=0,
xmax=40,
ymin=0,
ymax=100,
xmajorgrids,
ymajorgrids,
xlabel=Packet period (s),
ylabel=Packets (\%),
axis background/.style={fill=white}
]
\addplot [color=blue,solid,line width=1.5pt,mark=triangle,mark options={solid,draw=blue}]
  table[row sep=crcr]{%
0.5	6.1275\\
1	8.255\\
2	14.1225\\
5	35.2525\\
10	66.3075\\
15	92.6325\\
20	97.0075\\
25	99.3975\\
30	99.6425\\
35	99.7225\\
40	99.8725\\
};
\end{axis}
\end{tikzpicture}%
\vspace{-0.5em}
\caption{Successfully routed packets from any node vs. packet period.} 
\label{Fig:routing-pkt-received-vs-pkt-period}
\end{figure}
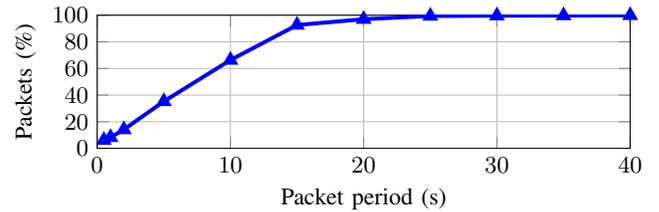

When a node is chosen as parent from several neighbors, 
or it must perform many retransmissions attempts,
its transmission queue can be fill up with unsent packets 
and further incoming packets are dropped.
Since the dropped incoming packets are never acknowledged, 
the link seems down for its neighbors.
This kind of congestion becomes worse when nodes operate at low data rates.
Although CTP is able to detect and surmount congested nodes, there is some performance limit. 
We have studied the average number of successfully routed packets 
from any node to the sink as a function of the packet periodicity.
Figure \ref{Fig:routing-pkt-received-vs-pkt-period} shows that the application layer should 
wait at least $25$ seconds between consecutive transmissions in order to successfully route $99.4\%$ of the packets to the sink, 
and it must wait $40$ seconds for routing $99.9\%$ of the packets.

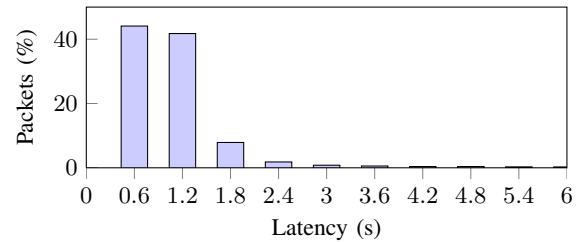
\begin{figure}[!ht]
\centering
%
%
%
\begin{tikzpicture}
\begin{axis}[
width=0.9\columnwidth,
height=0.42\columnwidth,
xlabel=Latency (s),
ylabel=Packets (\%),
xtick={0, 0.6, 1.2, 1.8, 2.4, 3, 3.6, 4.2, 4.8, 5.4, 6},
xmin=0,
xmax=6,
ymin=0,
ymax=50,
ybar,
xtick pos=left, 
ytick pos=left,
axis background/.style={fill=white}
]
\addplot [draw=black,fill=white!80!blue]
coordinates { 
(0.6, 44.1015)
(1.2, 41.7718)
(1.800, 7.8832)
(2.400, 1.8151)
(3.000, 0.8030)
(3.600, 0.5554)
(4.200, 0.3895)
(4.800, 0.3967)
(5.400, 0.2885)
(6.000, 0.2669)
(6.600, 0.2044)
(7.200, 0.1755)
(7.800, 0.1731)
(8.400, 0.1322)
};
\end{axis}
\end{tikzpicture}%
\vspace{-0.5em}
\caption{Latency histogram for a packet to be received by the sink.} 
\label{Fig:routing-ack-latency}
\end{figure}

Another important metric is the time that takes a packet from any node to be routed to the sink.
This latency depends on several issues, like how many hops are in the route, 
whether there are congested nodes that behave like bottlenecks, etc.
Figure \ref{Fig:routing-ack-latency} shows a latency histogram.
We have observed that, for this particular scenario,
$95\%$ of the received packets has latency lower than $2.5$ seconds,
$99\%$ lower than $9$ seconds 
and
$99.8\%$ lower than $25$ seconds.

In order to perform these simulations,
we extended the default \textbf{CtpTest} application 
such that the sink stores the packet number received and its source ID in a table.
This way, we can discharge duplicated packets (due to retransmission) 
and know whether each data packet has been received by the sink.
In addition, we increased the ACK wait delay to $1.7$ seconds (i.e., $5120$ bits) 
in the \textbf{CC2420Mac} header file to work better with our low bit rate.
We set the transmit power to $30$dBm
and used coherent receiver and channel coding,
so that we achieved good connectivity 
and every node could find a route to the sink.
Results have been averaged over 20 independent trials.

%
\subsection{Diffusion unicast transmission with ACK}
\label{ssec:diffusion-unicast-ack}
%

\begin{figure}[!ht]
\centering
\tikzset{
circle node/.style = {circle,inner sep=22pt,draw},
dot node/.style = {circle, draw, inner sep=5pt},
X node/.style = {fill=white, inner sep=1pt}
}

\begin{tikzpicture}[scale=0.9]
    \foreach \x in {0,...,6}
    \foreach \y in {0,...,2}
    {
    \fill (\x,\y) circle (2pt);
    }
    \node [circle node] at (3,1) {};
    
    \draw[-> , blue] (3, 1.3) -- (3, 1.7);   
    \draw[-> , blue] (3, 0.7) -- (3, 0.3);   
    \draw[-> , blue] (3.3, 1) -- (3.7, 1);   
    \draw[-> , blue] (2.7, 1) -- (2.3, 1);   
    
\end{tikzpicture}
\vspace{-0.5em}
\caption{Diffusion scenario.} 
\label{Fig:scenario-diffusion}
\end{figure}
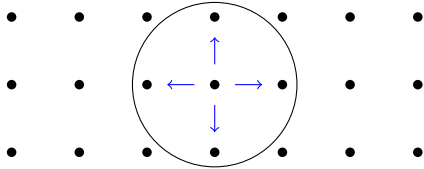

In this scenario, 
we consider the same network topology displayed in Figure \ref{Fig:scenario-routing}.
Nevertheless, instead of routing packets, 
here the nodes only communicate with their neighbors
(see Figure \ref{Fig:scenario-diffusion}),
as if they were performing an iterative distributed algorithm
(like those presented in companion paper \cite{Zazo2016}).

Similar to Sec \ref{ssec:routing-ack},
we use acknowledged unicast transmissions.
This means that every node has to figure out which nodes are its neighbors.
CTP offers an efficient mechanism for neighborhood discovery.
Hence, 
although we do not use routing in this scenario,
we have extended CTP so that this neighborhood discovery mechanism can be used for diffusion too.
This extension should be also useful for scenarios where nodes perform a distributed algorithm 
but also transmit data to the sink, combining autonomous and external monitoring.
We followed these steps to extend CTP:
\begin{enumerate}
\item Create an extra field in the routing packet that accounts for the address of the destination neighbor.
\item Extend \textbf{CtpForwardingEngine::encapsulatePacket} to account for the new field.
\item Include in \textbf{CtpForwardingEngine::event\_SubReceive\_receive}
the case where the node that receives the packet is the destination neighbor, 
so that it passes the packet to the application layer
instead of forwarding it to its parent.
\end{enumerate}

When using unicast transmissions, 
the nodes have to transmit and receive as many packets as neighbors per iteration.
In order to minimize congestion, 
we have modeled an asynchronous algorithm in which,
at every iteration,
every node transmits its packet to all its neighbors, 
waiting some time between unicast transmissions (unicast period).
Then, they wait some extra time--for 
allowing enough retries until the packets are acknowledged--before 
starting another iteration (contention period).
Figure \ref{Fig:channel_coefficient} shows 
the average percentage of successfully received packet 
vs. unicast period for different contention periods.
We have observed that for unicast period $5$ and contention period $100$ (seconds) 
the received packet rate is greater than $95$\%,
achieving $99$\% for $40$ seconds of unicast period.
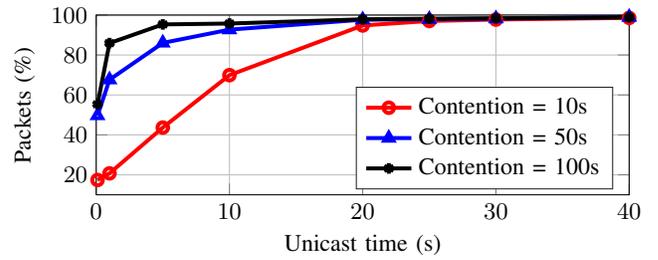
\begin{figure}[!ht]
\centering
%
%
\definecolor{mycolor1}{rgb}{0.00000,0.44700,0.74100}%
\definecolor{mycolor2}{rgb}{0.85000,0.32500,0.09800}%
\definecolor{mycolor3}{rgb}{0.92900,0.69400,0.12500}%
\definecolor{mycolor4}{rgb}{0.49400,0.18400,0.55600}%
\begin{tikzpicture}

\begin{axis}[%
width=0.8\columnwidth,
height=0.27\columnwidth,
scale only axis,
xmin=0,
xmax=40,
ymin=10,
ymax=100,
xmajorgrids,
ymajorgrids,
xlabel = Unicast time (s),
ylabel = Packets (\%),
axis background/.style={fill=white},
legend style={at={(0.97,0.03)},anchor=south east,legend cell align=left,align=left,draw=white!15!black}
]
\addplot [color=red,solid,line width=1.5pt,mark=o,mark options={solid,fill=red,draw=red}]
  table[row sep=crcr]{%
0.1	17.40325\\
1	20.7379\\
5	43.6653\\
10	69.9153\\
20	94.7903\\
25	97.0363\\
30	97.75805\\
40	98.59275\\
};
\addlegendentry{Contention = 10s};

\addplot [color=blue,solid,line width=1.5pt,mark=triangle,mark options={solid,draw=blue}]
  table[row sep=crcr]{%
0.1	49.69355\\
1	67.62905\\
5	86.0524\\
10	92.7016\\
20	97.68145\\
25	97.8992\\
30	98.20565\\
40	98.97175\\
};
\addlegendentry{Contention = 50s};

\addplot [color=black,solid,line width=1.5pt,mark=asterisk,mark options={solid,draw=black}]
  table[row sep=crcr]{%
0.1	55.31855\\
1	85.9476\\
5	95.31855\\
10	95.75805\\
20	97.88305\\
25	98.18145\\
30	98.38305\\
40	98.8508\\
};
\addlegendentry{Contention = 100s};


\end{axis}
\end{tikzpicture}%
\vspace{-0.5em}
\caption{Received packets vs. unicast period for multiple contention periods.} 
\label{Fig:diffusion-unicast}
\end{figure}

%
\subsection{Diffusion broadcast transmission (no ACK)}
\label{ssec:diffusion-broadcast}
%

In this section we also consider nodes transmitting to its neighbors 
over the same setting displayed by Figure \ref{Fig:scenario-diffusion}.
Nevertheless, here we suppose that the nodes broadcast only one packet for all its neighbors
(instead of transmitting one packet per neighbor).
This is achieved in Castalia by using the macro \textbf{BROADCAST\_NETWORK\_ADDRESS} 
as destination address.
We study two cases: \textit{synchronous} and \textit{asynchronous} transmissions.

\subsubsection{Synchronous transmissions}
\label{ssec:diffusion-broadcast-sync-transmission}

Every node transmits during its own time slot (i.e., TDMA).
Hence, 
no packet is received while the node is  transmitting,
meaning that there are neither interference, nor collisions.
The packet loss is only due to channel noise. 
In order to achieve this TDMA scheme we have used the application \textbf{connectivityMap}, setting the starting transmission time as follows:
\begin{align}
t_i^{\rm s} = P T_{\rm p}  N i
\label{eq:broadcast-synch}
\end{align}
where 
$t_i^{\rm s}$ denotes the starting time for node $i$,
$P=100$ is the number of packets 
(in a real algorithm this should be one, 
but here we want to average the channel noise),
$T_{\rm p} = 150$ms is the wait time between consecutive packets from the same node,
$N=21$ is the number of nodes in the network
and 
$i$ is the node identity.
Simulations show that we achieve $84.92$\% of successfully received packets.
This is a relatively high percentage with minimum number of transmitted packets 
(neither ACK, nor retries are necessary), 
but the cost of achieving accurate TDMA synchronization should be taken into account
when considering this option.

\subsubsection{Asynchronous transmissions}
\label{ssec:diffusion-broadcast-async-transmission}

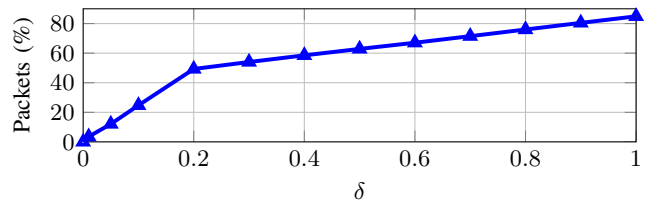
\begin{figure}[!ht]
\centering
%
%
\definecolor{mycolor1}{rgb}{0.00000,0.44700,0.74100}%
\begin{tikzpicture}

\begin{axis}[%
width=0.83\columnwidth,
height=0.2\columnwidth,
scale only axis,
xmin=0,
xmax=1,
ymin=0,
ymax=90,
xmajorgrids,
ymajorgrids,
xlabel=$\delta$,
ylabel=Packets (\%),
axis background/.style={fill=white}
]
\addplot [color=blue,solid,line width=1.5pt,mark=triangle,mark options={solid,draw=blue}]
  table[row sep=crcr]{%
0 	0\\
0.01 	3.53382\\
0.05		12.09297\\
0.1	24.72344\\
0.2	49.3\\
0.3	54.01563\\
0.4	58.51797\\
0.5	62.86641\\
0.6	67.05469\\
0.7	71.47891\\
0.8	75.98438\\
0.9	80.47578\\
1	84.89609\\
};
\end{axis}
\end{tikzpicture}%
\vspace{-0.5em}
\caption{Successfully received packets vs. asynchronous term $\delta$.} 
\label{Fig:diffusion-broadcast-async}
\end{figure}

There is no coordination among nodes.
Hence, the packet loss is due to noise, interference and collisions.
We have parametrized asynchronous transmissions 
by adding a term $\delta \in [0,1]$ to \eqref{eq:broadcast-synch}: 
\begin{align}
t_i^{\rm a} = \delta \cdot t_i^{\rm s}
\label{eq:broadcast-asynch}
\end{align}
When $\delta = 0$, 
we have $t_i = 0$ for all $i=1,\ldots, N$
so that all nodes transmit at the same time
(with exception of very small random clock drifts
that can be dismissed for short simulation time).
This makes all transmissions to collide
since 
no node can receive any packet while it is busy with its own transmission.
On the other hand, when $\delta = 1$, 
we recover the same result as for the perfect synchronization case studied in Sec. \ref{ssec:diffusion-broadcast-sync-transmission}.
Figure \ref{Fig:diffusion-broadcast-async} shows the average percentage of received packets
vs. the overlapping parameter $\delta$.

%
\section{Conclusion}
%

We presented preliminary simulation results of U-WSN in Castalia, 
a realistic, flexible and fast simulator.
We took real measurements of the underwater channel as starting point and proposed a simple linear path loss model.
We considered routing as well as diffusion scenarios and presented some results that are useful for designing centralized and distributed signal processing algorithms.
In order to perform the simulations, 
we had to extend Castalia, 
implementing the proposed path loss model, 
including coherent receiver and channel coding to the radio 
and adding diffusion capabilities to CTP.

%
\section*{Acknowledgment}
%

The authors would like to thank Juan Domingo Santana
Urbin (ULPGC) for all the help provided while making the underwater measurements
and to Gabriel Juanes and Raul Santana (PLOCAN) for setting up
the measurement testbed in the pier and into the sea.

%
\bibliographystyle{IEEEtran}
\bibliography{refs_sim_uwsn}
%

\end{document}